\documentclass[preprint,12pt]{elsarticle}
\usepackage{rotating}

\usepackage{amssymb}
\usepackage{amsmath}

\journal{Nuclear Physics A}

\begin{document}

\begin{frontmatter}



\title{Nuclear forces and nuclear excited states}

\author[label1]{Pawan Kumar}
\author[label2]{and Kanhaiya Jha}

 \affiliation[label1]{organization={GSI Helmholtzzentrum für Schwerionenforschung},
             city={Darmstadt},
             postcode={64291},
             state={Hessen},
             country={Germany}}

 \affiliation[label2]{organization={Medi-Caps University},
             city={Indore},
             postcode={453331},
             state={Madhya Pradesh},
             country={India}}

%

\begin{abstract}
In this work, the role of the central, spin-force and tensor forces of two-nucleon interaction in building the first 2$^+$ and 4$^+$ states of 20 sd-shell nuclei is studied. Calculations are performed within the framework of the nuclear shell model. It is shown that the central force predominantly contributes to the excitation energy. While the spin-orbit and tensor forces contribute relatively less, their roles are noted crucial in a few cases, such as in $^{22}$O. Further, it is demonstrated that the Hamiltonian projection approach should be preferred for rightly assessing the contributions of each force to the excitation energy. 
\end{abstract}




\end{frontmatter}



\section{Introduction}
\label{sec1}

Since the beginning of the twenty first century, the evolution of shell structure, also known as shell evolution, has remained one of the most discussed topics in the nuclear structure physics \cite{sorlin2008nuclear, otsuka2020evolution}. Along the technical developments in the experiments, it has attracted attention of scientists worldwide to the nucleon-nucleon interactions. In a novel manner, Otsuka and his collaborators demonstrated that the single-particle energy gaps evolve due to the empirical Gaussian central force and the $\pi + \rho$ mesons exchange tensor force acting between protons and neutrons \citep{otsuka2005evolution, otsuka2010novel}. In a complementary way, Umeya et al. \citep{umeya2004triplet, umeya2006single, umeya2016roles}, Smirnova et al. \citep{smirnova2010shell, smirnova2012nuclear}, and  Kumar et al. \citep{kumar2019proton, kumar2019quasi, kumar2023proton} have also explored the role of effective central, spin-orbit and tensor force components of shell model interaction in shell evolution. This studies are based on the unique method knows as spin-tensor decomposition \citep{kirson1973spin, klingenbeck1977central}, which allows the breakdown of shell effective interaction into its central and non-central components.

The role of spin-tensor decomposition has also remained vital in developing a comprehensive understanding of nucleon-nucleon interaction. It qualitatively demonstrate that the differences that arise in the intrinsic structure of shell model interaction after treating the many-body correlations \citep{wang2015revisiting,wang2014systematic,brown1988semi}. It has  been recently presented as a valuable tool for treating the discrepancy in the shell model interactions \citep{jha2020way, jha2020modification}. 

In the present work, we aim to employ spin-tensor decomposition to investigate the role of the central and non-central forces in constructing the low-energy spectrum. There are very few studies in the literature with this particular interest \citep{klingenbeck1977central, wang2015revisiting, kenji1980spin}. This might be due to fact that things becomes tricky when determining the excitation energy because of the involvement of diagonalization of Hamiltonian. As the shell model interactions involve non-vanishing central and non-central forces, it is fair to assume that the sum of the excitation energy values determined using them individually would not be equal to the excitation energy value obtained using the total force Hamiltonian. Hence, one should do the calculations with the complete force first and separate the contribution of central and non-central force later.
   
A condition must be met in order to use the spin-tensor decomposition, i.e.,  the model space must have the spin-orbit partners $j_>$ and $j_<$ corresponding to the quantum number $l$ \citep{kirson1973spin}. With this condition in the mind, we have focused on the sd-shell, in which the spin-orbit partners $d_{5/2}$ and $d_{3/2}$ with $l$ = 2, and its own spin-orbit partner $s_{1/2}$ with $l$ = 0 are present. I have reported the results for the first 2$^+$ and 4$^+$ states of 20 sd-nuclei in the present work. 

It should be noted that the present work complements the study by Wang et al. \citep{wang2015revisiting}, which demonstrates the effects of central and non-central forces on the excitation energy of the 2$^+$ state of O isotopes. However, the method used by the authors to compute the results differs from ours. The authors performed the diagonalization of the Hamiltonian of one force or a sum of two forces at a time and reported the results. In contrast, I performed the diagonalization of the complete Hamiltonian first and then determined the contribution of different forces following the projection approach. On a very positive note, this raises an important question how much the results vary between these two approaches. This has been briefly discussed later in the work.

This article is organized as follows. The details of the shell model interaction, the spin-tensor decomposition and the strategy followed for calculations are given in Section \ref{sec2}. The results are presented and discussed in Section \ref{sec3}. The summary is provided in Section \ref{sec4}.

\section{Theoretical Framework}
\label{sec2}
Among the many available shell model interactions for the sd-shell, we have performed calculations using USDB interaction \citep{brown2006new}. This interaction has been derived from the fit of the Bonn-A potential based renormalized G-matrix interaction to the 608 experimental data point, and till now, has remained the best choice in the sd-space for studying the spectroscopic properties of nuclei.

\begin{figure}
\centering
\includegraphics[height = 12cm, width=15cm, trim={1.0cm 5.5cm 5.0cm 1cm}]{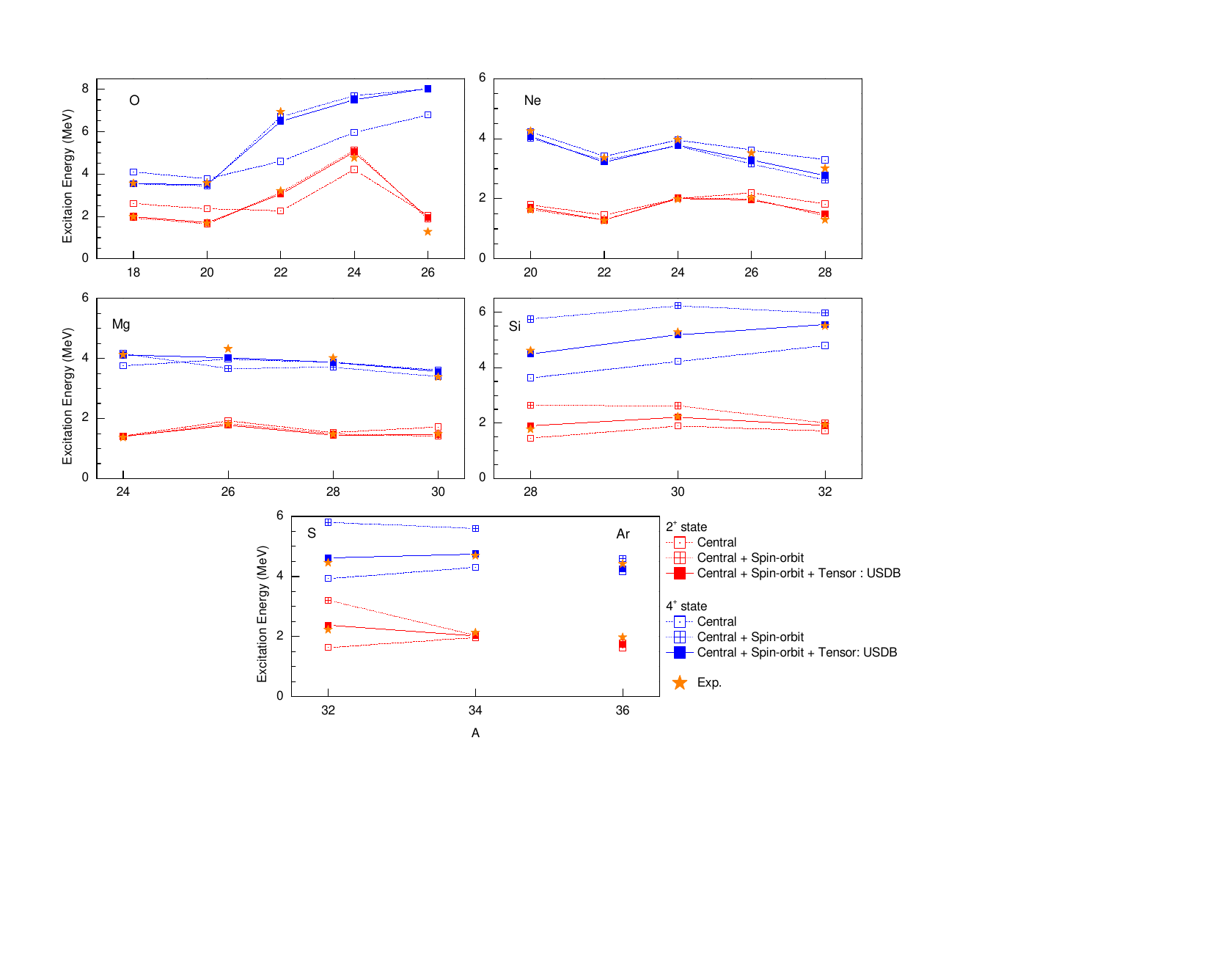}
\caption{Contribution of the central, spin-orbit and tensor forces of USDB interaction to the excitation energy of the 2$^+$ and 4$^+$ states of sd-shell nuclei. Experimental data is taken from the latest compilation available at NNDC \cite{NNDC2024}.}
\label{F1}
\end{figure}

In spin-tensor decomposition, basically the spin properties of nucleons are exploited. Since the nucleons are fermions with spin $1/2$, the interaction between two nucleons can be written as the linear sum of the scalar product of configuration space operator $Q$ and spin space operator $S$
\begin{equation}
V = \sum_{k=0}^2 V(k) = \sum_{k=0}^2 Q^k S^k ,
\label{E1}
\end{equation} where, rank $k$ = 0, 1 and 2 represent central, spin-orbit and tensor force, respectively. To obtain the matrix elements of these force in $jj$ coupling, which is followed in the shell model, the transform the $jj$ coupled state into to $LS$ coupled state is required. For the $LS$ coupled state, the matrix element $V(k)$ can be obtained from $V$ as
\begin{align}
\small
\langle (ab), LS: JM \mid V(k) \mid (cd), L' S': JM \rangle =  (2k+1) (-1)^J & \nonumber \\ 
\times 
\begin{Bmatrix}
L & S & J \\
S' & L' & k
\end{Bmatrix} 
\sum_{J'} (-1)^{J'} (2J'+1)  
\begin{Bmatrix}
L & S & J' \\
S' & L' & k
\end{Bmatrix}
\nonumber \\
\times \langle (ab), LS: J'M \mid V \mid (cd), L'S': JM \rangle
\label{E2}
\end{align}
where, $a$ is shorthand notation for the quantum numbers $n_a$ and $l_a$.

\begin{sidewaysfigure}
\includegraphics[width= 1.15\columnwidth, trim = {2.2cm 0cm 0cm 0cm}]
{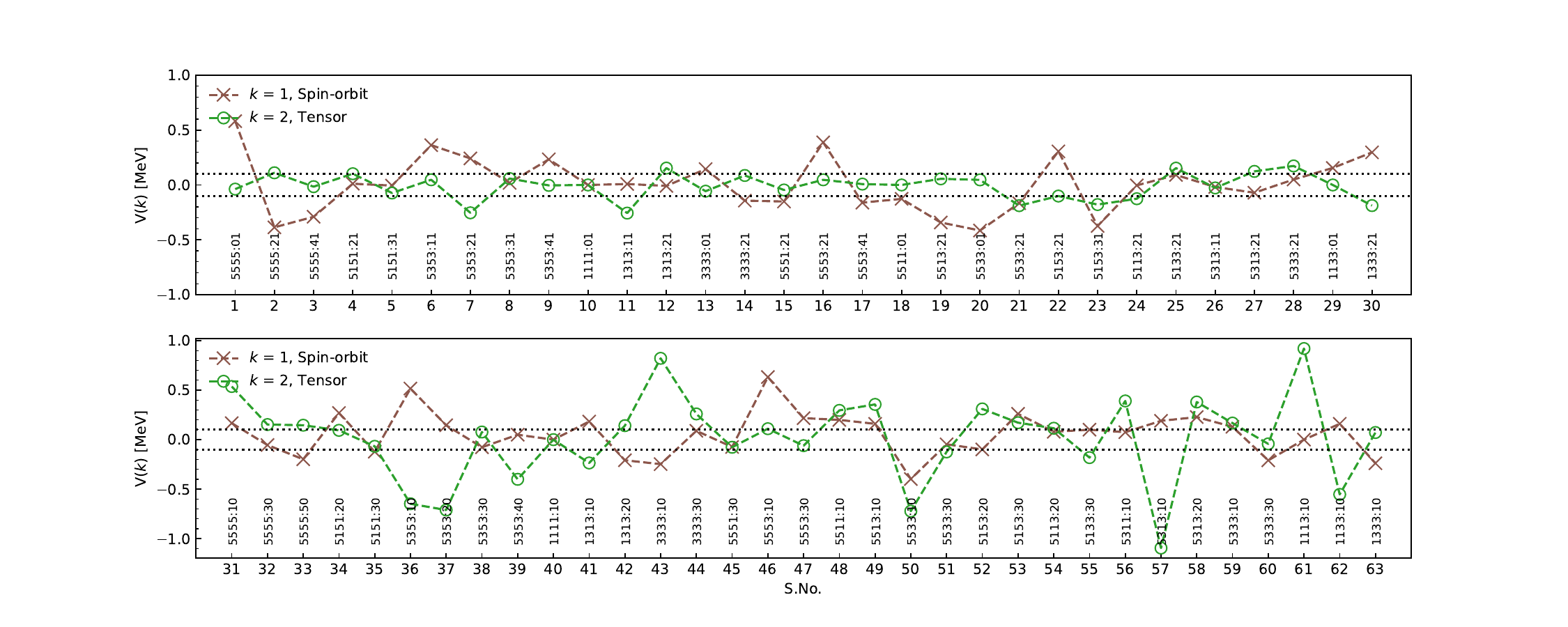}
\caption{Spin-orbit and tensor force matrix elements ($2j_{a}2j_{b}2j_{c}2j_{d}: JT$) of USDB interaction.}
\label{F2}
\end{sidewaysfigure}
 
\begin{figure*}
\includegraphics[height = 13cm, width=15.2cm, trim={3cm 2cm 0cm 0cm}]{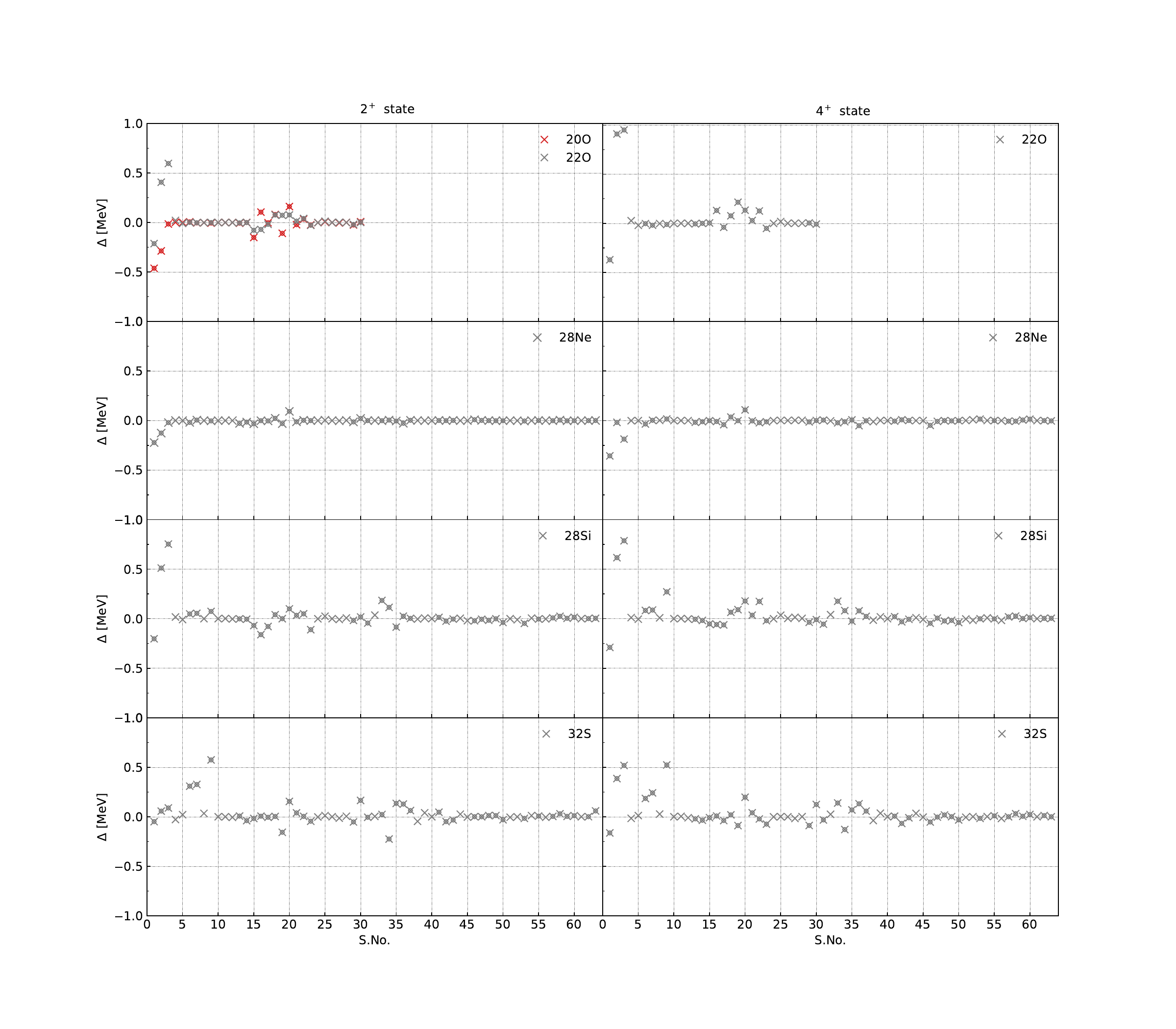}
\caption{Contribution of the spin-orbit force matrix elements to the 2$^+$ and 4$^+$ states of sd–shell nuclei. S.No at \textit{x}-axis follows the same numbering of matrix elements given in Fig.\ref{F2}. Symbol with open circle at the center of the cross refers to the matrix elements with magnitude exceeding 0.1 MeV. See text for details.}
\label{F3}
\end{figure*}

The large scale shell model calculations were carried out using the ANTOINE code \citep{caurier2005shell}. The initial calculations followed standard practice: the many-body Hamiltonian was constructed using the complete two-body force and diagonalized using the Lanczos method. Once the eigenvectors of the states were obtained, the Hamiltonian constructed using the central force and the central plus spin-orbit force was projected on them. As $[H_{i}, H]$ = 0 with $i$ = central and central plus spin-orbit force, the eigenvalues obtained through the Hamiltonian projection directly provide insights into the contribution of central and spin-orbit force to the excitation energy. It is important to note that no diagonalization was performed in the latter step.

\section{Results and Discussion}
\label{sec3}
The experiments have demonstrated that the oxygen isotopic chain contains two semi-magic nuclei, namely $^{24}$O and $^{26}$O \cite{thirolf2000spectroscopy, tshoo2012n}. In fig.~\ref{F1}, it can be seen that central force calculations do not justify the semi-magic title for $^{22}$O as the excitation energy of its 2$^{+}$ state is nearly degenerate with respect to that of $^{18}$O and $^{20}$O.  At $^{24}$O, the central force calculations show a sharp increase in E(2$^+$), a characteristic feature of semi-magic nuclei. When considering the spin-orbit force in the calculations, the E(2$^+$) of $^{18}$O and $^{20}$O decreases, whereas the E(2$^+$) of $^{22}$O increases. This restores the semi-magic behavior for the later. For the 4$^+$ state, the effect of spin-orbit force is similar to that in the 2$^+$ state. The tensor force is observed to be unimportant for both states. 

\begin{figure}
\includegraphics[height = 6cm, width=15.3cm, trim={3cm 0cm 0cm 0cm}]{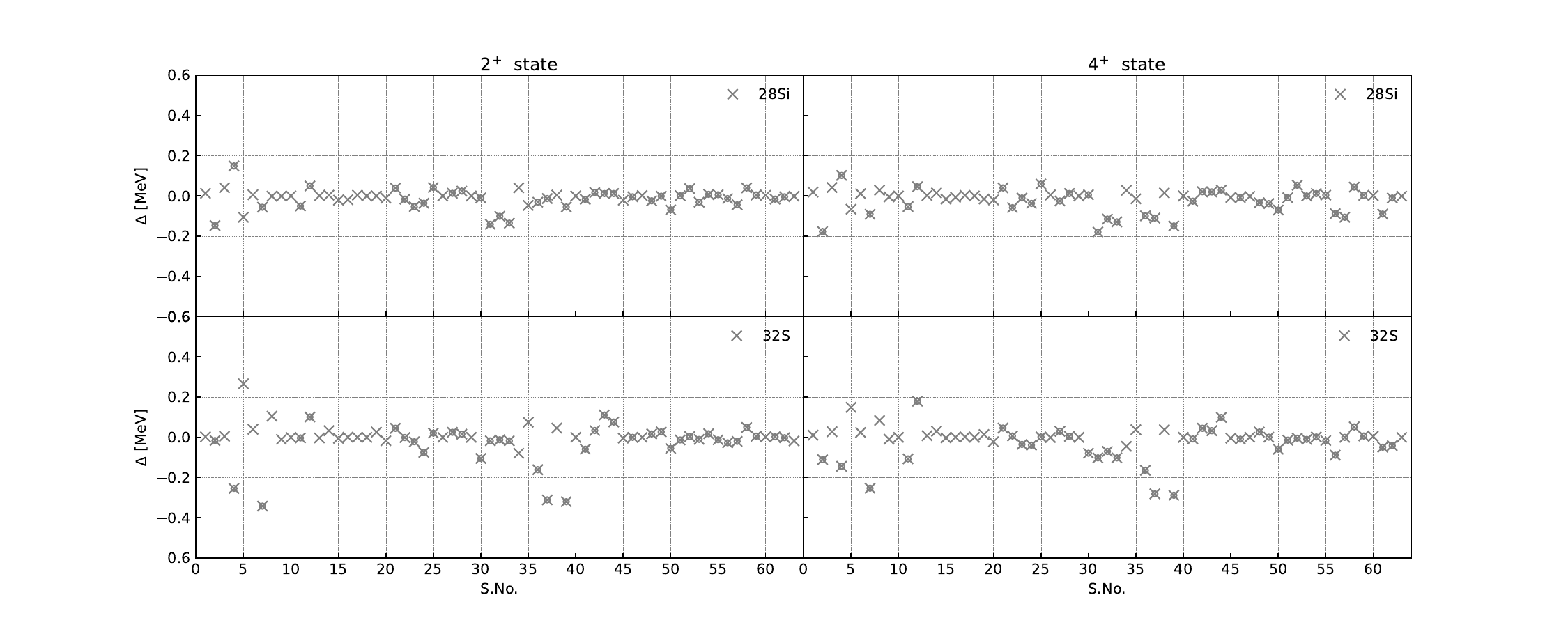}
\caption{Same as Fig. \ref{F3} but with tensor force matrix elements.}
\label{F4}
\end{figure}

\begin{figure}
\includegraphics[height = 6cm, width=15.3cm, trim={3cm 0cm 0cm 0cm}]{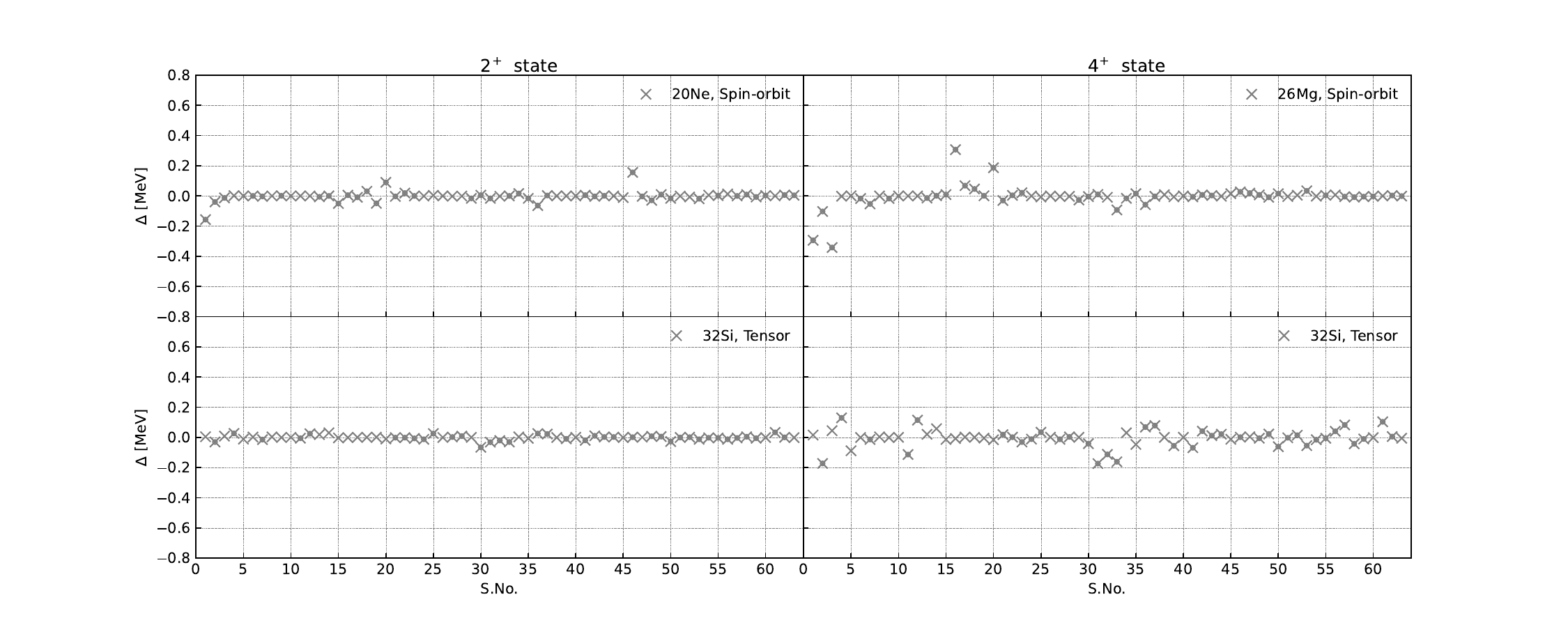}
\caption{Same as Fig. \ref{F3} but with both spin-orbit and tensor forces matrix elements}
\label{F5}
\end{figure}

\begin{figure}
\includegraphics[height = 18.5cm, width=15.2cm, trim={2cm 3cm 0cm 0cm}]{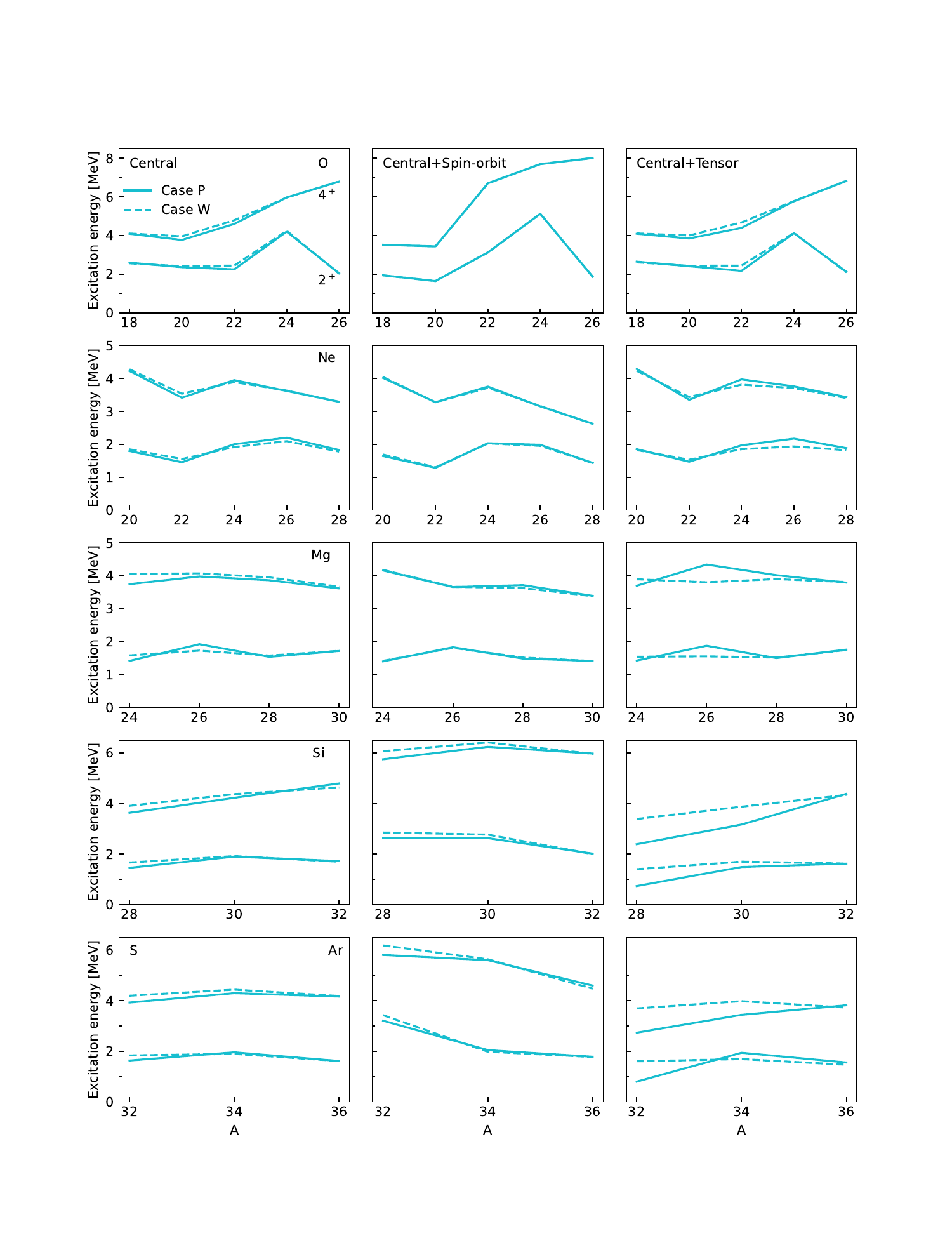}
\caption{Excitation energy of 2$^+$ and 4$^+$ states with central, central plus spin-orbit and central plus tensor force. The results were obtained following the projection of Hamiltonian and the direct diagonalization of Hamiltonian. They are shown with labels P and W, respectively.  }
\label{F6}
\end{figure}

In Ne isotopes, much like in Oxygen isotopes, the tensor force does not play any crucial role. It is central force that primarily contributes to excitation energy. At $^{28}$Ne, the excitation energy of both states needs to be decreased, and the spin-orbit force fulfills this requirement.

In Mg isotopes, surprisingly both spin-orbit force and tensor force basically acts as non-effective force, rendering the central force to be the decisive force to determine the excitation energy.

The effect of non-central forces is intriguing in Si isotopes. While the central force predominately contributes to E(2$^+$) and E(4$^+$), it underestimates the experimental values. The spin-orbit force increase the excitation energy of both states of $^{28,30}$Si, and 4$^+$ state of $^{32}$Si, which lead to overestimation. This situation is countered by the tensor force with the dropping effect and bring the calculated states closer to experimental states. 

The effects of spin-orbit and tensor force on the 2$^+$ and 4$^+$ states of $^{32}$S and the 4$^+$ state of $^{34}$S are the same as those observed in Si isotopes. On a smaller scale, both the spin-orbit and tensor forces similarly affect the 4$^+$ state of $^{36}$Ar. However, they counterpoise each other with almost equal strength which results in no net effect of non-central force on the state. In case of the 2$^+$ state of $^{34}$S and $^{36}$Ar, the individually effects of spin-orbit and tensor force are inconsequential. This allows the central force to completely determine the excitation energy. 

It becomes clear from Fig.\ref{F1} that the spin-orbit force and tensor force significantly influence the states in a few cases, while their influence is minimal in others. Figure \ref{F2} shows that more than half of the spin-orbit and tensor force matrix elements have magnitudes greater than 0.1 MeV. This raises following points: 1) which of these matrix elements affect the states the most, and 2) when the effects on the states are minimal, do these matrix elements act against each other or are their contributions trivial? To investigate these points, we have calculated the contribution of the spin-orbit and tensor force two body matrix elements to the excitation energy. This is done by projecting a Hamiltonian that comprises the whole central force (central and spin-orbit forces) plus one spin-orbit force (tensor force) matrix element onto the state vectors obtained with the total USDB interaction. To carry a qualitative discussion and avoid repetition, we have presented only a few cases below. Complete results can be provided on contact.

The results with the spin-orbit force are shown in Fig.~\ref{F4}. It can be seen that the 5555:01 ($2j_{a}2j_{b}2j_{c}2j_{d}: JT$), 5555:21 and 5555:41 matrix elements are crucial for restoring the semi-magic character of $^{22}$O. The first two matrix elements lower the 2$^{+}$ state in $^{20}$O whereas the first and the last matrix elements raise the same state in $^{22}$O. Further, the 5555:21 and 5555:41 matrix elements are necessary to elevate the 4$^{+}$ state in $^{22}$O. The 5555:01 matrix element, along with 5555:21 and 5555:41 matrix elements, lower the 2$^{+}$ and 4$^{+}$ states, respectively in $^{28}$Ne. The 5555:21 and 5555:41 matrix elements raise both 2$^{+}$ and 4$^{+}$ states in $^{28}$Si. The sets of three matrix elements 5353:11, 5353:21, 5353:41, and 5555:21, 5555:41, 5353:41 are useful to raise the 2$^{+}$ and 4$^{+}$ states in $^{32}$S.

In Fig.~\ref{F4}, it can be seen that no single or pair of tensor force matrix elements distinctively contribute to lowering the 2$^{+}$ and 4$^{+}$ states in $^{28}$Si. Instead, it is the collective result of several matrix elements that causes this effect. In $^{32}$S, the 5151:21, 5353:21, 5353:20 and 5353:40 matrix elements mainly acts against the spin-orbit force and lowers the states.

In Fig.~\ref{F5}, four cases related to minor changes in the states are shown. It can be observed that the spin-orbit and tensor force matrix elements trivially contribute to the 2$^+$ state of $^{20}$Ne and $^{32}$Si. To the 4$^{+}$ state of $^{26}$Mg, the 5555:01, 5555:41, 5553:21 and 5533:01 matrix elements notably contribute, but, with opposite sign. A similar situation occur in the 4$^{+}$ state of $^{26}$Mg where many matrix elements contributes with opposing effects.

In Fig.~\ref{F6}, a comparison is shown between the results obtained using the projection of the Hamiltonian (as discussed earlier in this work) and the direct diagonalization of the Hamiltonian (as used by Wang et al. \citep{wang2015revisiting}). Calculations were performed with the central, central plus spin-orbit, and central plus tensor forces. It can be seen that both methods give similar results for the central and central plus spin-orbit forces. However, they show discrepancies for the central plus tensor force. For example, in $^{28}$Si, the central plus tensor force predicts the 4$^{+}$ state at 2.38 MeV, whereas the diagonalization of the central plus tensor force Hamiltonian results in the same state being about 1 MeV higher. This significant difference highlights the importance of employing the projection method.
 
Solely focusing on oxygen isotopes, it can be seen that the tensor force has a minimal effect in both sets of calculations. Further, the contributions from the central and spin-orbit forces are similar. This led Wang et al. \citep{wang2015revisiting} and us to reach the same conclusion for oxygen isotopes. Authors also demonstrated that the spin-orbit force is crucial for achieving the two-nucleon unbound character of $^{26}$O. The same is noted in our calculations.

\section{Summary}
\label{sec4}

In the present work, the contribution of central, spin-orbit and tensor force to the first 2$^+$ and 4$^+$ states of sd-nuclei was studied within the framework of the nuclear shell model. To construct the many-body Hamiltonian, the phenomenological USDB interaction was considered. The decomposition of this nucleon-nucleon interaction into its internal two-body force structure was performed using spin-tensor decomposition. Results showed that the central force predominately contributes to the excitation energy. The spin-orbit and tensor forces contributed relatively less, but their contributions were effective in some cases. For example, the spin-orbit force is crucial in restoring the semi-magic character of $^{22}$O. 

Further, the results showed that the direct diagonalization of a smaller part of Hamiltonian may give unreliable results. Therefore, the method involving projection of smaller Hamiltonian should be preferred to obtain true results.     

\section*{Acknowledgments}
We would like to thank the IT department of GSI, Germany for providing the High-Performance Computing facilities for the calculations. Some of the computations were done when P.K. working at GSI under project number 448588010, funded by the Deutsche Forschungsgemeinschaft (DFG, German Research Foundation).

\bibliographystyle{elsarticle-num} 
\bibliography{references}
\end{document}